\documentclass[a4paper]{jpconf}
\usepackage{graphicx}
\usepackage{epstopdf}
\usepackage{subfigure}
\usepackage{textcomp}

\def\bkR{{\rm I\kern-.17em R}}
\def\bkC{{\rm \kern.24em \vrule width.05em height1.4ex depth-.05ex \kern-.26em C}}

\def\be{\beta}

\def\frac#1#2{{\textstyle{{#1}\over {#2}}}}
\def\laq{\raise 0.4 ex \hbox{$<$}\kern -0.8 em\lower 0.62 ex\hbox{$\sim$}}
\def\gaq{\raise 0.4 ex \hbox{$>$}\kern -0.7 em\lower 0.62 ex\hbox{$\sim$}}

\def\be{\begin{equation}}
\def\ee{\end{equation}}
\def\ba{\begin{eqnarray}}
\def\ea{\end{eqnarray}}

\def\dalemb#1#2{{\vbox{\hrule height.#2pt
        \hbox{\vrule width.#2pt height#1pt \kern#1pt \vrule width.#2pt}
        \hrule height.#2pt}}}

\def\dalemb#1#2{{\vbox{\hrule height.#2pt
        \hbox{\vrule width.#2pt height#1pt \kern#1pt \vrule width.#2pt}
        \hrule height.#2pt}}}

\def\gtorder{\mathrel{\raise.3ex\hbox{$>$}\mkern-14mu
             \lower0.6ex\hbox{$\sim$}}}
\def\ltorder{\mathrel{\raise.3ex\hbox{$<$}\mkern-14mu
             \lower0.6ex\hbox{$\sim$}}}

\begin{document}

\rightline{DF/IST-7.2008}
\rightline{December 2008}

\title{\bf Noncommutative Quantum Cosmology\footnote{Based on a talk presented by CB at DICE 2008, Castiglioncello, 22nd - 26th September 2008, Italy.}}

\author{C Bastos$^{1,2}$, O Bertolami$^{1,2}$, N C Dias$^{3,4}$ and J N Prata$^{3,4} $}

\address{$^1$ Departamento de F\'\i sica, Instituto Superior T\'ecnico, Avenida Rovisco Pais 1, 1049-001 Lisboa, Portugal \\
  $^2$ Instituto de Plasmas e Fus\~ao Nuclear, Instituto Superior T\'ecnico, Avenida Rovisco Pais 1, 1049-001 Lisboa, Portugal \\
  $^3$ Departamento de Matem\'atica, Universidade Lus\'ofona de Humanidades e Tecnologias, Avenida Campo Grande, 376, 1749-024 Lisboa, Portugal \\
  $^4$ Grupo de F\'isica Matem\'atica, Universidade de Lisboa, Avenida Prof. Gama Pinto 2, 1649-003, Lisboa, Portugal}

\ead {cbastos@fisica.ist.utl.pt, orfeu@cosmos.ist.utl.pt, ncdias@mail.telepac.pt, joao.prata@mail.telepac.pt}

\begin{abstract}

{One presents a phase-space noncommutative extension of Quantum Cosmology in the context of a Kantowski-Sachs (KS) minisuperspace model. We obtain the Wheeler-DeWitt (WDW) equation for the noncommutative system through the ADM formalism and a suitable Seiberg-Witten map. The resulting WDW equation explicitly depends on the phase-space noncommutative parameters, $\theta$ and $\eta$. Numerical solutions of the noncommutative WDW equation are found and, interestingly, also bounds on the values of the noncommutative parameters. Moreover, we conclude that the noncommutativity in the momenta sector leads to a damped wave function implying that this type of noncommutativity can be relevant for a selection of possible initial states for the universe.}

\end{abstract}

\section{Introduction}

Noncommutative space-time is believed to be a fundamental ingredient of quantum gravity \cite{Connes} and arises in String Theory/M-Theory, where a noncommutative effective low-energy gauge theory when describing the excitations of open strings in the presence of a Neveu-Schwarz constant background field \cite{Seiberg}. These ideas have sparked the interest in noncommutative field theories (QFT)[see e.g. \cite{Szabo} and Refs. therein]. In what concerns the one-particle sector of QFT, several noncommutative versions of quantum mechanics have been discussed (\cite{Gamboa}-\cite{Bastos}). The recent measurement of the first two quantum states of the gravitational quantum well for ultra cold neutrons has led one to consider noncommutative extensions of the gravitational quantum well \cite{Bertolami1,Bertolami2}. In that work, it is concluded that momentum noncommutativity should be considered for the noncommutative problem to differ from the commutative one.

The most general canonical noncommutative quantum mechanics (NCQM) is characterized by an extension of the Heisenberg-Weyl algebra. In general, time is assumed as a commutative parameter and a $2d$-dimensional phase-space of operators with noncommuting position and momentum variables is considered. This extended Heisenberg-Weyl algebra is given by:
\be\label{eq1.1}
\left[\hat q_i, \hat q_j \right] = i\theta_{ij} \hspace{0.2 cm}, \hspace{0.2 cm} \left[\hat q_i, \hat p_j \right] = i \hbar \delta_{ij} \hspace{0.2 cm}, \hspace{0.2 cm} \left[\hat p_i, \hat p_j \right] = i \eta_{ij} \hspace{0.2 cm},  \hspace{0.2 cm} i,j=1, ... ,d
\ee
where $\eta_{ij}$ and $\theta_{ij}$ are antisymmetric real constant ($d \times d$) matrices and $\delta_{ij}$ is the identity matrix. One can relate this extended algebra with the standard Heisenberg-Weyl algebra:
\be\label{eq1.3}
\left[\hat R_i, \hat R_j \right] = 0 \hspace{0.2 cm}, \hspace{0.2 cm} \left[\hat R_i, \hat \Pi_j \right]= i \hbar \delta_{ij} \hspace{0.2 cm}, \hspace{0.2 cm} \left[\hat \Pi_i, \hat \Pi_j \right] = 0 \hspace{0.2 cm}, \hspace{0.2 cm} i,j= 1, ... ,d ~,
\ee
by a class of linear (non-canonical) transformations:
\be\label{eq1.4}
\hat q_i = \hat q_i \left(\hat R_j , \hat \Pi_j \right) \hspace{0.2 cm},\hspace{0.2 cm} \hat p_i = \hat p_i \left(\hat R_j , \hat \Pi_j \right)
\ee
which are often referred to as the Seiberg-Witten (SW) map \cite{Seiberg}. With these transformations, one is able to convert a noncommutative system into a modified commutative one, which is dependent on the noncommutative parameters and of the particular SW map. The usual Schr\"{o}dinger equation with a modified $\eta,\theta$-dependent Hamiltonian determines the dynamics of the system and the wave functions in the Hilbert space correspond to the states of the system. However, it should be stressed that the physical properties of the system such as expectation values, probabilities and eigenvalues of operators are independent of the chosen SW map \cite{Bastos}.

In this contribution we report on the results of a study of the phase space noncommutative KS minisuperspace model \cite{Bastos2}. The KS minisuperspace model has been previously examined in the context of noncommutativity in the configuration space \cite{Compean,Barbosa}. We regard our more general approach fairly natural given that noncommutativity is presumably a relevant feature of quantum gravity and it is expected that its effects ought be important in the early universe. To address this question we implement the canonical quantization procedure in the context of the phase space noncommutative KS minisuperspace model in order to obtain the corresponding WDW equation of the problem (see e.g. Refs. \cite{Hartle, Bertolami3} for general discussions). 

Aiming to solve the noncommutative WDW equation, one implements a constraint that commutes with the Hamiltonian (and is associated with a constant of motion of the classical problem) and that allows for turning a  second order partial differential equation into an second order ordinary differential equation that is then solved numerically. By examining the resulting physical solutions one is able to restrict the set of possible values of the noncommutative parameters. 

Furthermore, we conclude that momentum space noncommutativity leads to a richer structure of states for the early universe. In this case the fundamental solutions of the WDW equation (which are featureless oscillations for both the commutative and configuration noncommutative cases) display a damping behaviour. This suggests a criterion for selection of states for the early universe. 

\section{The cosmological model}

One considers a cosmological model described by the KS metric. In the Misner parametrization, the line element can be written as \cite{Ryan}
\be\label{eq2.1}
ds^2=-N^2dt^2+e^{2\sqrt{3}\beta}dr^2+e^{-2\sqrt{3}\beta}e^{-2\sqrt{3}\Omega}(d\theta^2+\sin^2{\theta}d\varphi^2)~,
\ee
where $\beta$ and $\Omega$ are the scale factors and $N$ is the lapse function. The presence of at least two scale factors is necessary in order to impose a noncommutative algebra. The Hamiltonian for this metric is obtained via the ADM formalism \cite{Ryan}:
\be\label{eq2.1a}
H=N{\cal H}=Ne^{\sqrt{3}\beta+2\sqrt{3}\Omega}\left[-{P_{\Omega}^2\over24}+{P_{\beta}^2\over24}-2e^{-2\sqrt{3}\Omega}\right]~,
\ee
where $P_{\Omega}$ and $P_{\beta}$ are the canonical momenta conjugate to $\Omega$ and $\beta$, respectively. Given that the results are all gauge independent, one chooses $N=24e^{-\sqrt{3}\beta-2\sqrt{3}\Omega}$. This choice corresponds to a particular gauge choice, which is only motivated by technical simplicity. At quantum level, the treatment is manifestly covariant as the lapse function does not enter at all in the formalism. In the next sections, this will be shown explicitly. The classical and the quantum formulations will be considered separately.

\subsection{The Classical Model}

In the classical picture, the equations of motion for the four variables $\Omega$, $\beta$, $P_{\Omega}$ and $P_{\beta}$ can be obtained from the Poisson's bracket algebra. The commutative and the configuration space noncommuativity cases were discussed previously in Refs. \cite{Compean,Barbosa}. Analytical solutions to classical equations of motion were obtained for the two cases.

Here, one considers the most general noncommutative extension that can be obtained by imposing a noncommutative relation between the two scale factors, $\Omega$ and $\beta$, and between the two canonical momenta, $P_{\Omega}$ and $P_{\beta}$:
\be\label{eq2.10}
\left\{\Omega,P_{\Omega}\right\}=1\hspace{0.1 cm},\hspace{0.1 cm}\left\{\beta,P_{\beta}\right\}=1\hspace{0.1 cm},\hspace{0.1 cm}\left\{\Omega,\beta\right\}=\theta\hspace{0.1 cm},\hspace{0.1 cm}\left\{P_{\Omega},P_{\beta}\right\}=\eta~.
\ee
In the constrained hypersurface
\be\label{eq2.3a1}
{\cal H}\approx 0~,
\ee
the classical equations of motion for the noncommutative system are given by
\ba\label{eq2.11}
&&\dot{\Omega}=-2P_{\Omega}~,\hspace{0.5cm}(a)\nonumber\\
&&\dot{P_{\Omega}}=2\eta P_{\beta}-96\sqrt{3}e^{-2\sqrt{3}\Omega}~,\hspace{0.5cm}(b)\nonumber\\
&&\dot{\beta}=2P_{\beta}-96\sqrt{3}\theta e^{-2\sqrt{3}\Omega}~,\hspace{0.5cm}(c)\nonumber\\
&&\dot{P_{\beta}}=2\eta P_{\Omega}~.\hspace{0.5cm}(d)
\ea
Due to the entanglement among the four variables, an analytical solution is hard to find. However, it is possible to obtain a numerical solution to this system as well as a constant of motion from Eqs. (\ref{eq2.11}a) and (\ref{eq2.11}d):
\be\label{eq3.1}
\dot{P_{\beta}}=-\eta(-2P_{\Omega})=-\eta\dot{\Omega}\Rightarrow P_{\beta}+\eta\Omega=C~;
\ee
this will play an important role in solving the phase space noncommutative WDW equation.

\subsection{The Quantum Model}

Here and henceforth, one assumes a system of units where $c=\hbar=G=1$, that is the Planck units. Regarding the noncommutative parameters, $\theta$ and $\eta$, as an intrinsic feature of quantum gravity, they are expected to be of order one in Planck units.

By canonical quantization of the classical Hamiltonian constraint, Eq. (\ref{eq2.3a1}), one obtains the commutative WDW equation for the wave function of the universe. For the simplest ordering of operators,
\be\label{eq2.2}
\exp{(\sqrt{3} \hat{\beta}+2\sqrt{3} \hat{\Omega})}\left[- \hat P^2_{\Omega}+ \hat P^2_{\beta}-48e^{-2\sqrt{3} \hat{\Omega}}\right]\psi(\Omega,\beta)=0~.
\ee
where $\hat P_{\Omega}=-i \frac{\partial }{\partial \Omega}$, $\hat P_{\beta}=-i \frac{\partial }{\partial \beta}$ are the fundamental momentum conjugate operators to $\hat{\Omega} = \Omega$ and $\hat{\beta} = \beta$, respectively. Notice that, as it is usual in Quantum Cosmology, Eq.(\ref{eq2.2}) depends on the prescribed factor ordering. One chooses the simplest factor ordering, which has already been studied in Refs. \cite{Compean,Barbosa}. This allows for a direct comparison with the results in the literature.

For the commutative case, the solutions of Eq. (\ref{eq2.2}) are \cite{Compean},
\be\label{eq2.3}
\psi^{\pm}_{\nu}(\Omega,\beta)=e^{\pm i\nu\sqrt{3}\beta}K_{i\nu}(4e^{-\sqrt{3}\Omega})~,
\ee
where $K_{i\nu}$ are modified Bessel functions.

One requires that the coordinates and the canonical momenta do not commute. Thus, the extended Heisenberg algebra is,
\be\label{eq2.4}
\left[\hat{\Omega}, \hat{\beta} \right]=i\theta\hspace{0.2 cm},\hspace{0.2 cm}\left[\hat P_{\Omega}, \hat P_{\beta}\right]=i\eta\hspace{0.2 cm},\hspace{0.2 cm}\left[\hat{\Omega}, \hat P_{\Omega}\right]=\left[\hat{\beta},\hat P_{\beta}\right]=i~.
\ee
To obtain a representation of this algebra (\ref{eq2.4}), one transforms it into the standard Heisenberg algebra through a SW map \cite{Bastos2}:
\ba\label{eq2.8}
\hat{\Omega} =\lambda \hat{\Omega}_{c}-{\theta\over2\lambda} \hat P_{\beta_c} \hspace{0.2cm} , \hspace{0.2cm} \hat{\beta} = \lambda \hat{\beta}_{c} + {\theta\over2\lambda} \hat P_{\Omega_c}~,\nonumber\\
\hat P_{\Omega}= \mu \hat P_{\Omega_c} + {\eta\over2\mu} \hat{\beta}_{c} \hspace{0.2cm} , \hspace{0.2cm} \hat P_{\beta}=\mu \hat P_{\beta_c}- {\eta\over2\mu} \hat{\Omega}_{c}~,
\ea
where the index $c$ denotes commutative variables, i.e. variables for which $\left[\hat{\Omega}_c, \hat{\beta}_c\right]=\left[\hat P_{\Omega_c}, \hat P_{\beta_c}\right]=0$ and $\left[\hat{\Omega}_c, \hat P_{\Omega_c}\right]=\left[\hat{\beta}_c, \hat P_{\beta_c}\right]=i$. This transformation can be inverted only if:
\be\label{eq3.1a}
\xi \equiv \theta \eta <1.
\ee
In this case, the inverse transformation reads:
\ba\label{eq3.2}
\hat{\Omega}_c={1\over\sqrt{1- \xi}}\left( \mu \hat{\Omega} + {\theta\over2\lambda} \hat P_{\beta}\right) \hspace{0.2cm} , \hspace{0.2cm} \hat{\beta}_c={1\over\sqrt{1-\xi}} \left( \mu \hat{\beta} -{\theta\over2\lambda} \hat P_{\Omega}\right)~,\nonumber\\
\hat P_{\Omega_c}={1\over\sqrt{1-\xi}} \left(\lambda \hat P_{\Omega}-{\eta\over2\mu} \hat{\beta} \right) \hspace{0.2cm} , \hspace{0.2cm} \hat P_{\beta_c}={1\over\sqrt{1-\xi}}\left( \lambda \hat P_{\beta}+{\eta\over2\mu} \hat{\Omega} \right)~.
\ea
Substituting the noncommutative variables, expressed in terms of the commutative ones, into the commutation relations (\ref{eq2.4}), one obtains a relation between the dimensionless constants $\lambda$ and $\mu$:
\be\label{eq2.8a}
\left(\lambda\mu\right)^2-\lambda\mu+{\xi\over4}=0\Leftrightarrow\lambda\mu={1+\sqrt{1-\xi}\over2}~.
\ee
Hence, through the transformation Eq. (\ref{eq2.8}), one may regard Eq. (\ref{eq2.4}) as an algebra of operators acting on the usual Hilbert space $L^2(\bkR^2)$. In this representation, the WDW Eq. (\ref{eq2.2}) is deformed into a modified second order partial differential equation, which exhibits an explicit dependence on the noncommutative parameters:
\be\label{eq2.9}
\left[-\left(-i \mu {\partial \over \partial {\Omega_c}}+{\eta\over2\mu}\beta_{c}\right)^2+\left(-i \mu {\partial \over \partial {\beta_c}}-{\eta\over2\mu}\Omega_c\right)^2-48\exp{\left[-2\sqrt{3}\left(\lambda\Omega_c+i{\theta\over2\lambda} {\partial \over \partial {\beta_c}} \right)\right]}\right]\psi(\Omega_c,\beta_c)=0~.
\ee
This equation is fairly complex and cannot be solved analytically. However, the noncommutative quantum version of the constant of motion Eq. (\ref{eq3.1}):
\be\label{eq3.3}
\hat{C}=\hat{P_{\beta}}+\eta\hat{\Omega}=\sqrt{1-\xi}\left(\mu \hat{P}_{\beta_c}+{\eta\over2\mu}\hat{\Omega}_c\right)
\ee
commutes with the noncommutative Hamiltonian constraint Eq. (\ref{eq2.9}), for the chosen operator ordering. This allows one to transform the partial differential Eq. (\ref{eq2.9}) into an ordinary differential equation, which can be then solved numerically \cite{Bastos2}. The results are presented in the next section.

\section{Solutions}

We consider now Eq. (\ref{eq2.9}) in detail. Defining $\hat{A}=\frac{\hat{C}}{\sqrt{1-\xi}}$ from Eq. (\ref{eq3.3}), it then follows that:
\be\label{eq3.4}
\mu \hat{P}_{\beta_c}+{\eta\over2\mu}\hat{\Omega}_c=\hat{A}~.
\ee

As already referred to, the noncommutative WDW Eq. (\ref{eq2.9}) is very complex and no analytical solution is likely to be found. The strategy is to solve it numerically transforming it into an ordinary differential equation. First, one verifies that the constant of motion, Eq. (\ref{eq3.1}), commutes with the Hamiltonian in the constrained space of states:
\be\label{eq3.7}
\left[\hat{P}_{\beta}+\eta\hat{\Omega},\hat{H}\right]=\left[\hat{P}_{\beta}+\eta\hat{\Omega},-\hat{P}^2_{\Omega}+\hat{P}^2_{\beta}-48e^{-2\sqrt{3}\hat{\Omega}}\right]=0~.
\ee
Then, one looks for solutions of Eq. (\ref{eq2.9}) that are simultaneous eigenstates of the Hamiltonian and of the constraint Eq. (\ref{eq3.4}). Let $\psi_a(\Omega_c,\beta_c)$ be an eigenstate of the operator Eq. (\ref{eq3.4}) with a real eigenvalue, $a \in \bkR$, then:
\be\label{eq3.8}
\left(-i\mu{\partial\over\partial\beta_c}+{\eta\over2\mu}\Omega_c\right)\psi_a(\Omega_c,\beta_c)=a \psi_a(\Omega_c,\beta_c)~.
\ee
This equation admits the solution:
\be\label{eq3.9}
\psi_a(\Omega_c,\beta_c)=\Re(\Omega_c)\exp{\left[{i\over\mu}\left(a-{\eta\over2\mu}\Omega_c\right)\beta_c\right]}~.
\ee
Substituting the wave function (\ref{eq3.9}) into Eq. (\ref{eq2.9}) one obtains
\ba\label{eq3.10}
&&{\mu}^2\left[{\Re''\over\Re}-i{\eta\over\mu^2}{\Re'\over\Re}\beta_c-{\eta^2\over4\mu^4}\beta_c^2\right]+i\eta\left[{\Re'\over\Re}-i{\eta\over2\mu^2}\beta_c\right]\beta_c-{\eta^2\over4\mu^2}\beta_c^2+\left(a-{\eta\over2\mu}\Omega_c\right)^2-\nonumber\\
&&-{\eta\over\mu}\left(a-{\eta\over2\mu}\Omega_c\right)\Omega_c+{\eta^2\over4\mu^2}\Omega_c^2-48\exp{\left[-2\sqrt{3}\left(\lambda\Omega_c-{\theta\over2\lambda\mu}\left(a-{\eta\over2\mu}\Omega_c\right)\right)\right]}=0~,
\ea
where $\Re'\equiv\frac{d \Re}{ d \Omega_c}$. After some algebraic manipulations, one finally gets
\be\label{eq3.11}
\mu^2\Re''+\left(\eta{\Omega_c\over\mu}-a\right)^2\Re-48\exp{\left[-2\sqrt{3}{\Omega_c\over\mu}+{\sqrt{3}\theta\over\lambda\mu}a\right]}\Re=0~.
\ee
Performing the following change of variables,
\be\label{eq3.12}
z={\Omega_c\over\mu}\hspace{0.2 cm}\rightarrow\hspace{0.2 cm}{d\over dz}=\mu{d\over d\Omega_c}~,
\ee
it finally yields for $\phi(z)\equiv\Re(\Omega_c(z))$
\be\label{eq3.13}
\phi''(z)+\left(\eta z-a\right)^2\phi(z)-48\exp{[-2\sqrt{3}z+{\sqrt{3}\theta\over\lambda\mu}a]}\phi(z)=0~.
\ee
This second order ordinary differential equation can be solved numerically. Its solutions depend on the eigenvalue $a$ and on the noncommutative parameters $\theta$ and $\eta$.

\begin{figure}
\begin{center}
\subfigure[ ~$\theta=\eta=0$ and $a=0.4$]{\includegraphics[scale=0.7]{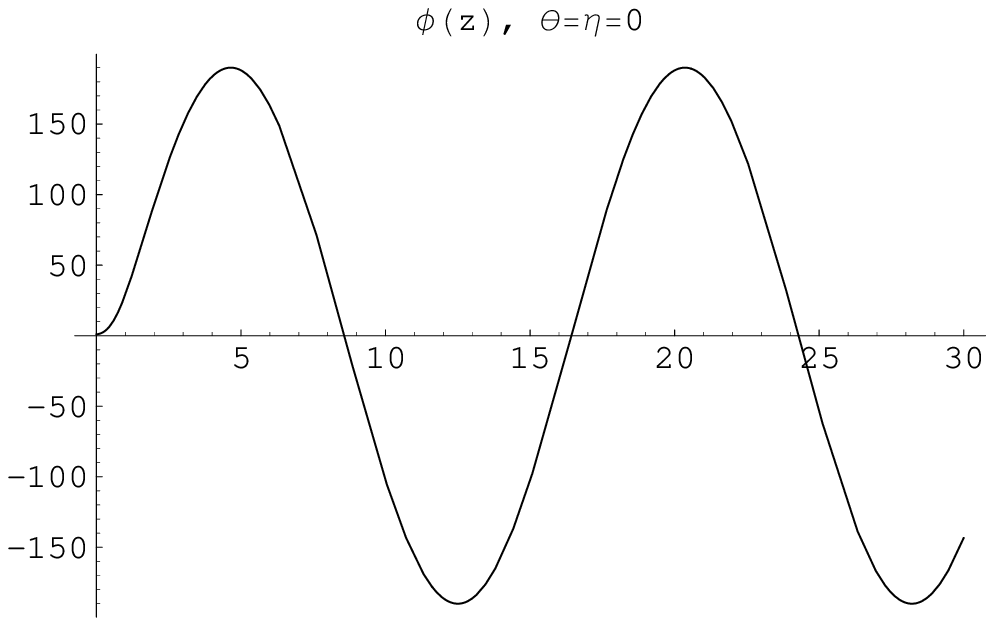}}
\subfigure[ ~$\theta=5$, $\eta=0$ and $a=0.4$]{\includegraphics[scale=0.7]{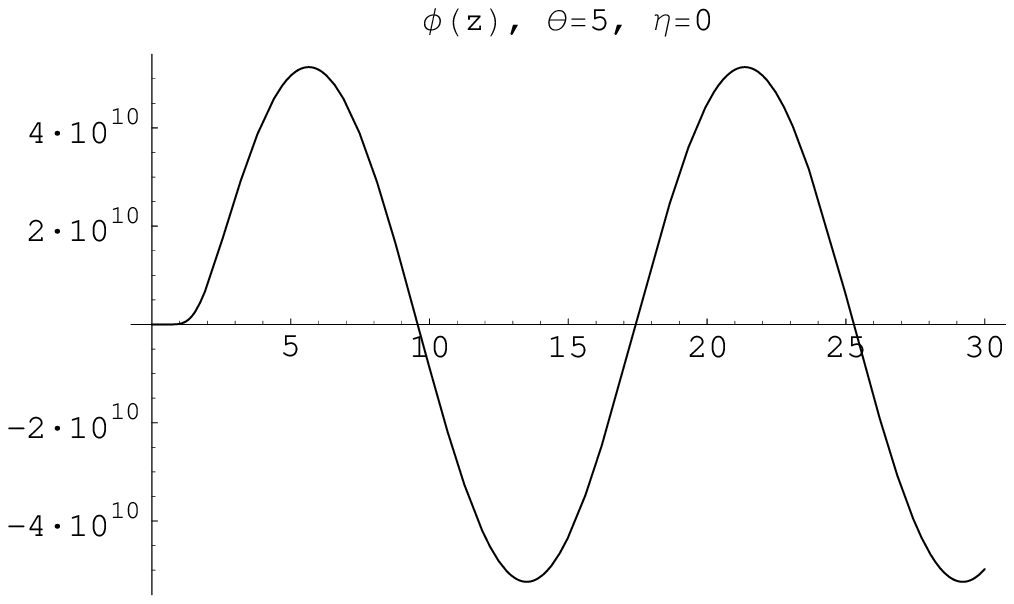}}
\subfigure[ ~$\theta=0$, $\eta=0.1$ and $a=0.565$]{\includegraphics[scale=0.7]{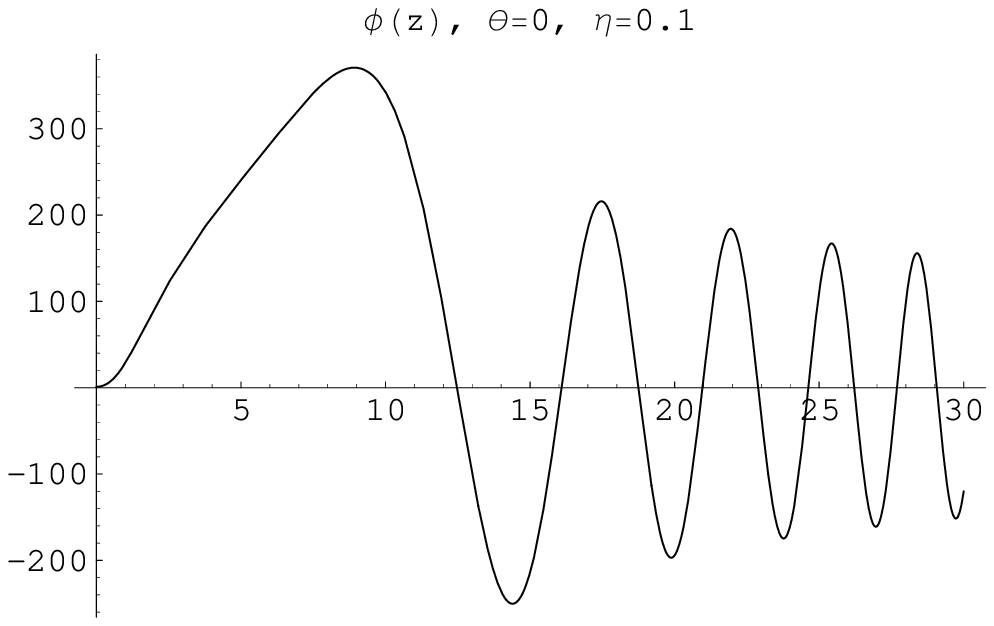}}
\subfigure[ ~$\theta=5$, $\eta=0.1$ and $a=0.799$]{\includegraphics[scale=0.7]{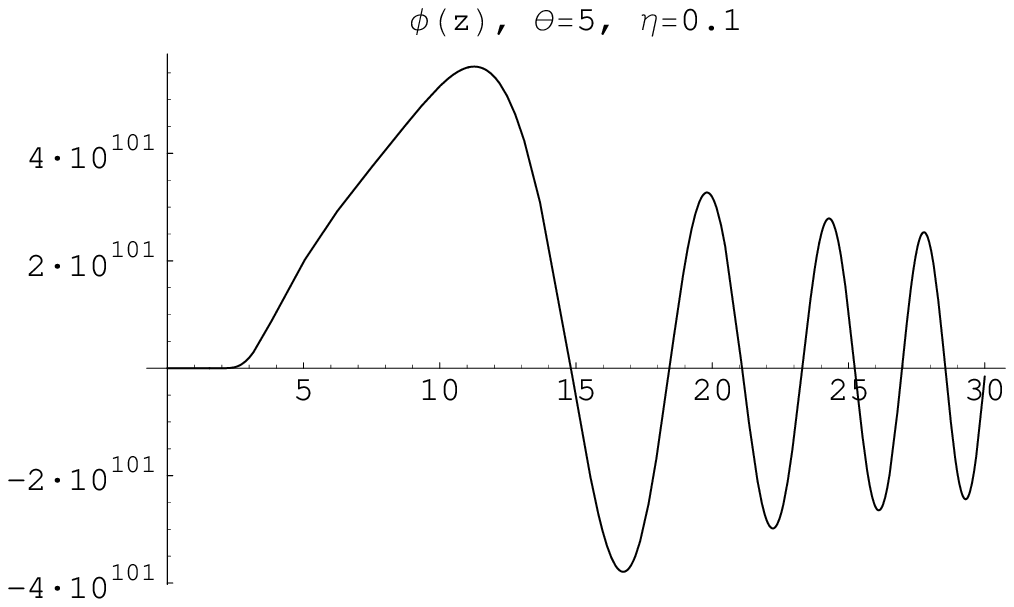}}
\caption{Representation of the numerical solutions of Eq. (\ref{eq3.13}) for different values to the noncommutative parameters. In the four plots $P_{\beta}(0)=0.4$ and $\Omega(0)=1.65$.}
\label{funcaodeonda}
\end{center}
\end{figure}

Our problem involves with four initial conditions: $\Omega(0)$, $\beta(0)$, $P_{\Omega}(0)$ and $P_{\beta(0)}$. With the exception of $\beta(0)$, all the others are related to each other due to the constraint Eq. (\ref{eq2.3a1}). Thus, if we choose some numerical values for $P_{\beta(0)}$ and $P_{\Omega}(0)$, one immediately obtains a value for $\Omega(0)$. $\beta(0)$ is an independent initial condition.

In Fig. \ref{funcaodeonda} we present the numerical solutions of Eq. (\ref{eq3.13}) for particular sets of values for $a, \theta$ and $\eta$. The eigenvalue $a$ was taken to be $a=\frac{C}{\sqrt{1-\theta\eta}}$ and is determined through Eq. (\ref{eq3.1}) from the classical values $P_{\beta}(0)$ and $\Omega(0)$ used to generate the solutions of Eqs. (\ref{eq2.11}). These classical values are fairly typical, and once again they were borrowed from the previous studies of the noncommutative KS cosmological model \cite{Barbosa}, so as to allow for comparison with previous results. However, one finds that under variations of the relevant parameters, the qualitative behaviour of the obtained wave function is not significantly changed.

Indeed after a thorough analysis of the results, one observes that the qualitative features of the solutions displayed in Fig. \ref{funcaodeonda} is unchanged for a rather broad range of values for $\theta,\eta$ and $a$ \cite{Bastos2}. The choice $\theta=5$ is fairly typical in what concerns the properties of the wave function. Furthermore, it is consistent with the point of view that the noncommutative parameters should be of order one close to the fundamental quantum gravity scale. The summary of the main results of Ref. \cite{Bastos2} is the following:

\begin{enumerate}
\item For $\theta=5$, the wave function has a damping behaviour for $\eta$ in the range $0.05<\eta<0.12$;
\begin{itemize}
	\item For $\eta_c>0.12$  the wave function blows up, suggesting that it is an upper limit for momenta noncommutativity;
\end{itemize}

\item For $\theta>\eta$, varying $\theta$ affects the numerical values of $\phi(z)$, but its qualitative features remain unchanged. 
\begin{itemize}
	\item The lower limit for $\eta$ which exhibits a damping behaviour is around $\eta\sim0.05$ for all $\theta>\eta$. Clearly, higher $\eta$ values (c.f. Fig. \ref{funcaodeonda}) have a great influence on the wave function;
\end{itemize} 
\begin{itemize}
	\item For $\eta=0$ the wave function is essentially oscillatory;
\end{itemize}
\begin{itemize}
	\item For $0<\eta<0.05$, the wave function is actually amplified instead of exhibiting a damping behaviour;
\end{itemize}

\item For $\eta>\theta$, the damping behaviour of the wave function is harder to observe as the wave function does not blow up only for certain values of $\theta$, in particular when $\eta\in[1,2]$;
\begin{itemize}
	\item For $\eta\geq3$ there are no possible ranges for $\theta$ for which the wave function is well defined;
\end{itemize}

\item For large $z$ the qualitative behaviour of the wave function is analogous to the one depicted in Fig. \ref{funcaodeonda} for $z\leq30$.

\end{enumerate}

The criterion to determine bounds for the noncommutative parameters is based on the existence of well defined smooth solutions of the WDW equation. As discussed, solutions do not exist for arbitrary values of $\theta$ and $\eta$. Notice that it is the noncommutativity introduced in the momenta space that affects the behaviour of the wave function. It is interesting to observe that this kind of noncommutativity turned oscillatory functions of the commutative and noncommutative in configuration space cases into damped wave functions. This novel feature is a welcome property as it introduces structure into the wave function, suggesting a natural selection of states for the quantum cosmological model and thus for the set of initial conditions of the classical cosmological model. 

\section{Conclusions}

In this contribution we have reviewed the effects of a phase space noncommutativity on the minisuperspace KS quantum cosmological model. Through the analysis of the noncommutative system, one finds a classical constant of motion that allows for a numerical solution of the noncommutative WDW equation. One clearly sees that the quantum model is affected by the noncommutativity on the momenta sector, that is the wave function presents a damping behaviour for growing values of the $\Omega$ variable. Thus, the wave function is more peaked for small values of $\Omega$, which is a rather interesting and a new portrait of the quantum behaviour for the very early universe.

\ack

\noindent The work of CB is supported by Funda\c{c}\~{a}o para a Ci\^{e}ncia e a Tecnologia (FCT) under the fellowship SFRH/BD/24058/2005. The work of OB is partially supported by the FCT project No. POCTI/FP/63916/2005. The work of NCD and JNP was partially supported by Grant No. POCTI/0208/2003 and PTDC/MAT/69635/2006 of the FCT.


\end{document}